\documentclass[fleqn,10pt]{wlscirep}
\usepackage[utf8]{inputenc}
\usepackage[T1]{fontenc}

\usepackage{graphicx,color}
\usepackage{amsmath,amssymb}
\usepackage{verbatim}
\usepackage{ulem}
\usepackage{mathptmx}
\usepackage{dsfont}
\usepackage{ulem}

\def\spvec#1{\left(\vcenter{\halign{\hfil$##$\hfil\cr \spvecA#1;;}}\right)}
\def\spvecA#1;{\if;#1;\else #1\cr \expandafter \spvecA \fi}

\title{Employing machine learning for theory validation and identification of experimental conditions in laser-plasma physics}

\author[1,2,3,*]{A.~Gonoskov}
\author[4]{E.~Wallin}
\author[5]{A.~Polovinkin}
\author[3]{I.~Meyerov}
\affil[1]{Chalmers University of Technology, SE-41296 Gothenburg, Sweden}
\affil[2]{Institute of Applied Physics, RAS, Nizhny Novgorod 603950, Russia}
\affil[3]{Lobachevsky State University of Nizhni Novgorod, Nizhny Novgorod 603950, Russia}
\affil[4]{Department of Physics, Ume\aa~University, SE-90187 Ume\aa, Sweden}
\affil[5]{Adv Stat \& Machine Learning, LTD, Intel, USA}

\affil[*]{arkady.gonoskov@chalmers.se}

\begin{abstract}
	The validation of a theory is commonly based on appealing to clearly distinguishable and describable features in properly reduced experimental data, while the use of ab-initio simulation for interpreting experimental data typically requires complete knowledge about initial conditions and parameters. We here apply the methodology of using machine learning for overcoming these natural limitations. We outline some basic universal ideas and show how we can use them to resolve long-standing theoretical and experimental difficulties in the problem of high-intensity laser-plasma interactions. In particular we show how an artificial neural network can "read" features imprinted in laser-plasma harmonic spectra that are currently analysed with spectral interferometry.
\end{abstract}
\begin{document}

\flushbottom
\maketitle

\thispagestyle{empty}

\section*{Introduction}

Over the last few years the use of machine learning opened up new vistas in many areas of physics, including plasma physics \cite{spears.pop.2018}, condensed-matter physics \cite{carrasquilla.natp.2017, deng.prb.2017}, quantum physics \cite{krenn.prl.2016, carleo.sci.2017, briecker.sr.2017, chng.prx.2017}, thermodynamics \cite{torlai.prb.2016}, quantum chemistry \cite{li.qc.2016}, particle physics \cite{baldi.natc.2014} and many others. Recent examples include applications for magnetic confinement fusion \cite{cannas.nf.2007, vega.fed.2013, rea.fst.2018}, inertial confinement fusion \cite{humbird.arxiv.2017, nora.sadm.2017, peterson.pop.2017}, discovery of phase transitions \cite{wang.prb.2016, nieuwenburg.natp.2017, hu.pre.2017}, closure of turbulence models \cite{tracey.aiaa.2015}, representation of quantum states \cite{gao.natc.2017, glasser.prx.2018}, galaxy classification \cite{huertas-company.aj.2018} and orbital stability \cite{lam.mnras.2018}. One of the origins of this progress is the possibility of processing large sets of data and drawing conclusions based on features that admit no straightforward description and assessment with human languages. In this way, some natural human limitations can be overcome, making machine learning be a useful tool that works in a fruitful synergy with traditional approaches in theoretical and experimental physics.

In this paper we consider the opportunities of using machine learning for solving long-standing problems in laser-plasma physics. We discuss the possibility of using autonomous recognition of difficult-to-qualify features in the data of real or numerical experiments for validating and advancing phenomenological models as well as for reconstructing experimental conditions. Using a phenomenological model for laser-plasma high-harmonic generation, we train an artificial neural network (NN) to reconstruct various parameters based on the recognition of unspecified features in the harmonic spectra. The NN then "identifies" the learned features in the spectra obtained with ab-initio simulations, which we use to mimic real experiments. In this way we can reconstruct the parameters of experiments or determine the most appropriate values for free parameters of incomplete theories. This can also be used to determine the validity regions of different models. It is important that this approach can be applied in case of inaccurate or intrinsically incomplete knowledge about the experimental conditions, i.e. in cases when performing a particular ab-initio simulation is not possible. In such a way this approach can provide new routes for experiments and new insights for theory and model development. 

For the sake of completeness, we start from the discussion of basic ideas, using the Galton Board \cite{galtonboard} as an illustrative example. We then provide a proof-of-principle demonstration of the use of this methodology for the outlined problem in the field of laser-plasma interactions. 

\section*{Methods}
 
Typically, the validation of a theory is reduced to the experimental observation of some clearly describable feature, such as an observable physical value, its certain dependency on some parameters, a peak in some distribution, etc. These kinds of features are conventionally used to claim for the agreement of experimental and theoretical results.

One natural limitation of this conventional approach is that in the presentation of results we appeal to the consistency in terms of such clearly describable features and consequently do this mostly for features in one, two or three-dimensional sets of data. The essential part of many studies is finding ways of reducing and transforming raw data sets into the forms that expose such describable, indicative features. Of course, there have been developed numerous techniques and approaches, such as statistical and spectral analysis as well as various algebraic transformations. However, this toolbox is inevitably limited and in many cases the solution of the outlined problem requires insightful analysis and development of the theory and experiment with, sometimes, manual search for such a feature in large sets of data. 
%\cite{guest.arxiv.2018}

Another common consequence of developing sophisticated comparison methodology is losing the lucidity in the relation between the compared features and the origins of the theory. In some cases, it is difficult to say whether the observed feature unambiguously indicate the correctness of the theory or if it is peculiar to a family of theories that are thus not disqualified by the experiment. For example, the selected feature can be a generic consequence of some conservation laws rather than of the main principle or assumption that has to be validated. In other words, it can be difficult to quantify in what sense and to what degree the theory is validated and what the alternative theories that are disqualified by the experiment are. 

One more limitation of such conventional methodologies is the fact that theories can be benchmarked against the data of experiments with sufficiently complete knowledge about the initial conditions and all important parameters. In some cases, the information is intrinsically incomplete that hampers the use of theories and ab-initio simulations.

In this paper we discuss and apply a methodology that overcomes the outlined limitations with the help of machine learning applied to the recognition of hardly describable features in outputs of ab-initio simulations or experimental data, even in case of essentially incomplete knowledge about the experimental conditions. We provide proof-of-principle demonstration of several essential capabilities of this approach:

\begin{enumerate}
	
	\item Comparison, validation or disqualification of theories in a lucid and quantitative way (as a function of position in parameter space);
	\item Completing theories through indirect measuring dependencies of free parameters, even in the parameter regions where they are not well-defined in terms of the first principles;
	\item Indirect measurement for determination of unknown experimental parameters;
	\item Identifying regions in a parameter space where certain ranges of experimental data carry unambiguous information about experimental conditions.
	
\end{enumerate}

Note that these approaches are not intended to replace traditional methodologies but to supplement them with methods of gaining knowledge that can either be exploited heuristically or used to conceive hypothesis and ideas for further theoretical and experimental studies.

Although many ideas might look very trivial and known, we start from a general discussion of the outlined approaches. To support the discussion we use the well-known Galton Board \cite{galtonboard} experiment as a simple example of an experiment and a theory. After that we demonstrate how the developed methods can be used for resolving long-standing questions in the physics of high-intensity laser-plasma interactions.

\subsection*{Validation of a theory}

Suppose we need to validate a theory $\mathds{A}$ using an experiment $\mathds{E}$. To formulate the problem in a more exact way we assume that we intend to compare theory $\mathds{A}$ with some alternative theory $\mathds{B}$ (or a family of alternative theories) which we intend to disqualify using experimental data. In our notations, both theories $\mathds{A}$ and $\mathds{B}$, as well as the experiment $\mathds{E}$ are denoted as some possibly non-linear and non-deterministic operators that act on a vector of initial conditions $\textbf{c}$  and give a vector of measurable quantities $\textbf{r}$. These vectors can represent a set of data of arbitrary composition and dimensionality. Suppose we carried out a sequence of experiments, then we can write:

\begin{eqnarray}\label{def1}
&& \textbf{r}_i^A = \mathds{A} \textbf{c}_i, \\
&& \textbf{r}_i^B = \mathds{B} \textbf{c}_i, \\
&& \textbf{r}_i^E = \mathds{E} \textbf{c}_i, 
\end{eqnarray}
where index $i$ enumerates the experiments. We admit that due to experimental imperfections the values $\textbf{r}_i^E$ are a subject of some unknown systematic or non-systematic distortions that hamper direct comparison of $\textbf{r}_i^E$ with $\textbf{r}_i^{A}$ and $\textbf{r}_i^{B}$. However, we assume that the measurable data $\textbf{r}$ contains some features that can appear in the results of either $\mathds{A}$ or $\mathds{B}$. These features can depend on the conditions $\textbf{c}$ in a complex way, which also should be reproduced by the appropriate theory (at least to some extent). Note, that in general both theories can be applicable in certain regions of the parameter space of vectors $\textbf{c}$ and this is something that we intend to determine.\\

To solve the problem we develop a unification theory $\mathds{U}$ that depends on at least one parameter $p$ and provides a smooth transition between the theories, for example, $\mathds{U}(p = 0) = \mathds{B}$, and $\mathds{U}(p = 1) = \mathds{A}$. In the most primitive case, this can be just a linear combination, i. e. $\mathds{U}(p) = p \mathds{A} + (1 - p) \mathds{B}$. However, it is better to form the smooth transition not in-between the final results of the theories, but between the essential principles or assumptions that provide the origin for the development of the theories. This is because, even if both theories fulfill basic conservation laws, their linear combination might not (for $0 < p < 1$). To avoid this we can extend the dimensionality of the parameter $p$, so that in this space there exist some route between the points corresponding to $\mathds{A}$ and $\mathds{B}$ so that at each point of this route all the conservation laws are fulfilled. With help of simple examples we will see further why it is important.\\

We can now apply the unified theory for various possible initial conditions and generate a sufficiently large set of pairs $\textbf{c}_k$ and $\textbf{r}_k^{U} = \mathds{U} \textbf{c}_k$ for various random values of $p$ and conditions $\textbf{c}$ ($k$ is the index running over the set). Next, we train a feed forward fully connected NN $\mathds{N}$ to learn how to reconstruct $p$ and $\textbf{c}$ from $\textbf{r}_k$, i. e. 
\begin{equation}
\left(p_k^N, \textbf{c}_k^N \right) = \mathds{N} \textbf{r}_k^U. 
\end{equation}
According to \cite{hornik.nn.1991} any continuous real-valued function can be approximated with such a NN for any given accuracy. Thus, we choose this type of NN as an approximation of unknown function that maps $\textbf{p}$ and $\textbf{c}$ to $\textbf{r}$. We can then apply the trained NN to the experimental data $\textbf{r}_i^E$. If the values of $\textbf{c}^E$ are systematically close to the reconstructed values $\textbf{c}^N$ in the whole or some certain region of parameter space, we can interpret this as if the NN "recognizes" some indicative features in this region of parameter space. In this region the reconstructed value $p$ can indicate the validity of one of the theories: a systematic tendency of $p$ to 1(0) indicates the validity of theory $\mathds{A}$($\mathds{B}$). This procedure can also show the transition between the applicability regions of theories explicitly, through plotting $p$ as a function of parameters $\textbf{c}$. 

It would be reasonable to ask: In what sense is the validity of a theory is demonstrated by this procedure? Of course, there exists a large variety of relations between the output value $\textbf{r}$ and the parameter $p$ and certainly not all of them are necessarily sensible in terms of physics. In other words, the NN can establish successful correlation between $p$ and some feature of little importance or complete irrelevance to the physics of the process. To avoid this, we train the NN to reproduce not only $p$, but a sufficiently large set of physically essential values $\textbf{c}$. This favours establishing correlations with some features that significantly depend on the initial conditions and, in this sense, have some physical meaning. Although a more rigorous analysis would be of interest, in this paper we focus on showing proof-of-principle examples that convincingly demonstrate the rationality of this concept.

We can outline one important peculiarity of the method: the procedure does not require a complete knowledge about the experimental parameters $\textbf{c}_i$.  We can use only the known components of vector $\textbf{c}$ to see whether the NN "recognizes" the features of importance or not.

As a proof-of-principle example we consider the Galton Board (GB), which is also known as a bean machine. We highlight that the Galton Board is chosen as a clear illustrative example, while one can certainly apply standard statistical methods for this problem. The reasons and some advantages of using machine learning will be discussed later in a separate section.

We use index $j$ to denote the final position of a bead that bypasses $n$ horizontal rows of pegs. Bouncing from each peg leads to equal probability of bypassing it from each of two sides. The probability of coming to the $j$-th positions is then given by
\begin{equation}
r_j^A =  \spvec{j;n} 0.5^j 0.5^{n - j} \approx \sqrt{\frac{2}{\pi n}}\exp\left(- \frac{2}{n}\left(j - \frac{n}{2}\right)^2\right),
\label{a_bm}
\end{equation}
where the approximation is the de Moivre-Laplace theorem applied under the assumption of $n \gg 1$. We will use this Gaussian distribution of limited applicability as theory $\mathds{A}$, which we intend to validate using experiments. As the experiment for this problem we will use the numerical implementation of Monte-Carlo method, with a limited number of beads. The theory $\mathds{A}$ will be validated relative to an alternative theory $\mathds{B}$ that suggests that the distribution is super-Gaussian:
\begin{equation}
r_j^B \sim \exp\left(- p_2\left(j - \frac{n}{2}\right)^4\right),
\end{equation}
We use this form as a particular example because it is difficult to describe and appeal to the difference between the Gaussian and super-Gaussian distribution without elaborating and applying additional data processing. We here deliberately do not apply any transformation to show the capability of the used approach. Note also that theory $\mathds{B}$ is formulated incompletely. The missing factor in front of the exponent can be determined using the normalization conditions. We here intentionally do not compute this factor to demonstrate how one can deal with incomplete theories.

We now define a unified theory $\mathds{U}$:
\begin{equation}
r_j^U = p_0 \sqrt{\frac{8}{\pi n}} \exp\left(-p_1\left(j - \frac{n}{2}\right)^2 - p_2 0.1 \left(j - \frac{n}{2}\right)^4\right).
\label{u_bm}
\end{equation}
We see that theory $\mathds{A}$ corresponds to $p_1 \neq 0$, $p_2 = 0$ while theory $\mathds{B}$ is characterized by $p_1 = 0$, $p_2 \neq 0$. (More precisely, theory $\mathds{A}$ corresponds to $\textbf{p} = \left(0.5, 2/n, 0\right)$, but this is not important for now.) Note, that there exist a value of $p_0$ that provides probability normalization. In other words, within the family of theories $\mathds{U}$ there exist theories of type $\mathds{B}$ that fulfill the probability normalization. As we will see, we can disqualify $\mathds{B}$ even without knowing this value of $p_0$.  

We now randomly generate values of $p_0$, $p_1$ and $p_2$ from 0 to 1, generate the result $r_j$ according to the unified theory $\mathds{U}$ and train the NN to reconstruct the generated vector $\textbf{p} = (p_0, p_1, p_2)$. In our implementation we used $n = 16$ and the result was sampled through 16 values $r_j$. For the proof-of-principle demonstration we used a rather small feed forward fully connected NN that contained four layers with the neuron numbers 16,~16,~16 and ~3, respectively. The three output neurons were associated with the components of the vector $\textbf{p}$. We used logistic sigmoid, squared error measure and stochastic gradient descent, which resulted in an accuracy of $\textbf{p}$ determination of the order of $10^{-3}$. 

Next we perform a number of numerical experiments using a random number generator to simulate a number of beads that pass through $n = 16$ layers of pegs. The number of beads reached each of the positions is normalized by the total number of beads to get the experimental values $r_j^E$. These distributions are then used as input for the NN, which reconstructs $\textbf{p}$ according to the learned unified theory.

\begin{figure}[t!]
	\centering\includegraphics[width=0.5\columnwidth]{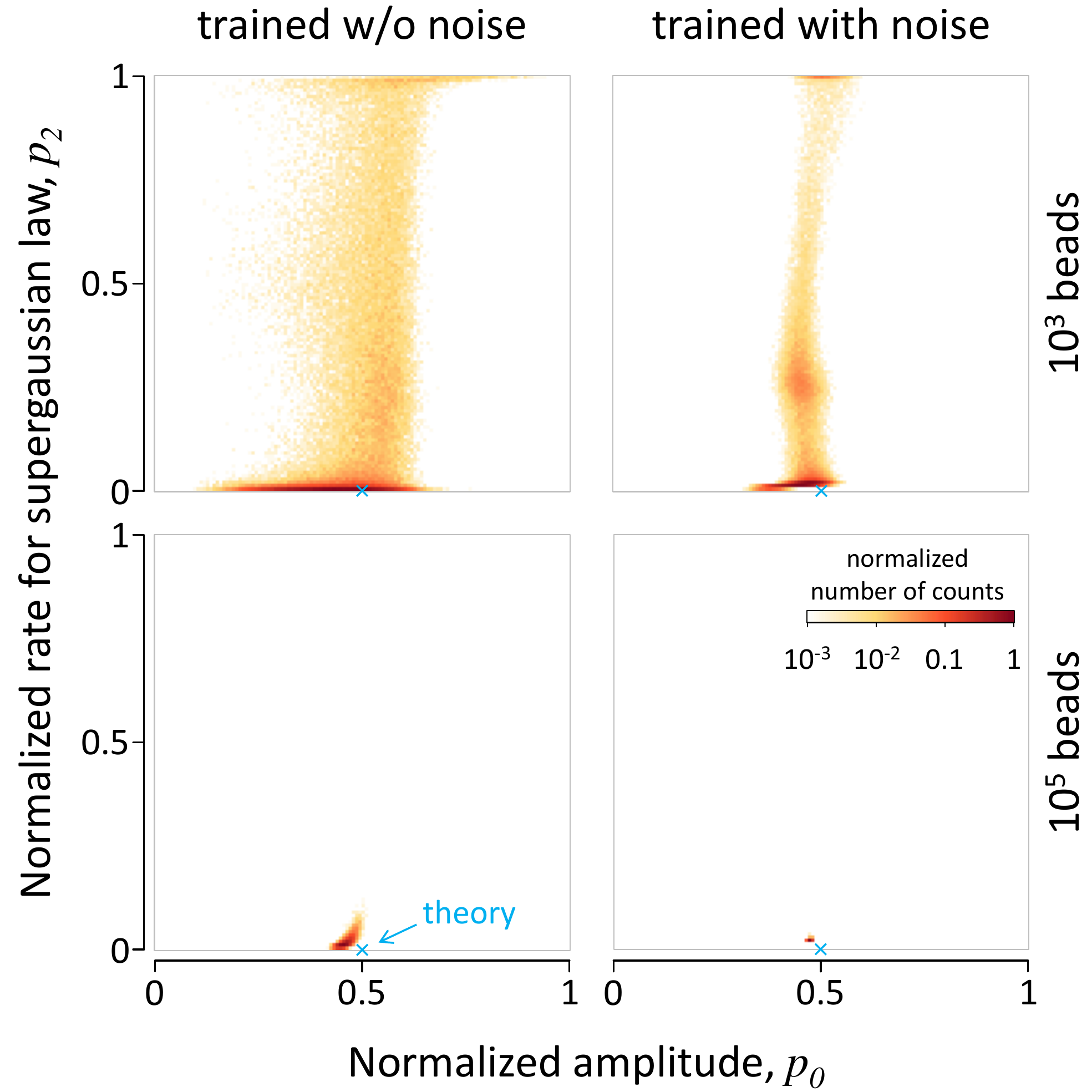}
	\caption{Validation of the theory for the Galton Board in the limit of large number of layers of pegs (Eq.~(\ref{a_bm})). The plots show the distribution of reconstructed values of parameters of the unified theory (Eq.~(\ref{u_bm})) by the NN that receives the distributions obtained with Monte-Carlo method using $10^3$ (upper panels) and $10^5$ (lower panels) beads. The tendency of $p_2$ to 0 indicates irrelevance of the super-Gaussian component and invalidity of such alternative theory. The right column shows the results of using a NN that is trained with additional noise that leads to higher tolerance to the experimental noise caused by smaller number of beads.}
	\label{bm_tv}
\end{figure} 

In Fig.~\ref{bm_tv}~(left column) we plot the distribution of reconstructed values on the plane of $p_0$ and $p_1$ for the number of beads equal to $10^3$ (upper panels) and $10^5$ (lower panels). As we see, in both cases the reconstructed values are localized mostly in the vicinity of the point related to the theory $\mathds{A}$, i.e. ($p_2 = 0$ and $p_1 = 0.5$). However, this is more obvious in the case of using the larger number of beads. This is not surprising because the NN was trained to reproduce values based on exact distribution without any stochastic deviations. To establish some tolerance of the NN to noise we train it using an artificial noise that we apply to the theoretical values before sending them to the input of the NN. For this purpose we multiply each value $r_j^U$ by a factor $(1 - 0.005 + 0.01r)$, where $r$ is random value from 0 to 1. As we can see from the comparison of the panels in Fig.~\ref{bm_tv}, this results in a better localization of the distribution around the expected point (precise analysis shows that this improvement appears for both $p_2$ and $p_0$). This result demonstrates that this approach can be used to retrieve more efficiently the information from the experimental results with noise.

We can note that the distributions of reconstructed values are centered at a point that is close to, but notably different from the point predicted by the theory. This is an indication of the fact that the theory is valid in the limit $n \gg 1$, while we here use a large, but finite value, $n = 16$. In other words, we see the closest fitting of experimental data in the framework of theories described by the unified theory $\mathds{U}$. This observation leads to the next opportunity that we identify and describe in the next subsection. 

\subsection*{Completing a theory}

Suppose we have an incomplete theory $\mathds{A}(\alpha)$ that contains some free parameter $\alpha$ (or several parameters). Alternatively we can have a hypothesis that some theoretical approach can be applied under the appropriate choice of a parameter that we were not able to determine. We can use the methodology described in the previous subsection to determine both the possibility and the appropriate values in such cases.

We generate a sequence of theoretical results for various values of parameters $\textbf{c}$ and values of $\alpha$ in the appropriate range: 
\begin{equation}
\textbf{r}_i^A = \mathds{A}(\alpha_i) \textbf{c}_i,
\end{equation}
and train a NN to reconstruct the values of $\textbf{c}$ and $\alpha$. We can then vary the experimental conditions $\textbf{c}$ and send each result to the input of the trained NN. The agreement of the reconstructed values to the known experimental parameters $\textbf{c}$ would indicate that the NN "relates" some features in the input to the ones peculiar to the theory $\mathds{A}(\alpha)$. If such agreement is not observed, this procedure does not indicate weather such a theory can be applied or not. Alternatively, if the agreement for $\textbf{c}$ is observed, but the values of $\alpha$ are not exposing any systematic tendency, we can conclude that there is no appropriate choice for free parameter $\alpha$.

However, if, in some region of parameters, the reconstructed values of $\textbf{c}$ are systematically close to the known values in the experiment, this procedure can show the systematic dependency of $\alpha$ on the parameters $\textbf{c}$.

One can approximate the obtained dependencies and use the approximations as a heuristic way to complete the theory. However, this can also provide a hint for further development of a theory in a deductive way. For example, the possibility of applying certain assumptions or phenomenological model can become clear and be validated rigorously.

Finally, we would like to highlight that this procedure can be applied to determine the experimental conditions that are not known or even not measurable. This procedure can thus be treated as indirect measurement. 

\begin{figure}[t!]
	\centering\includegraphics[width=0.4\columnwidth]{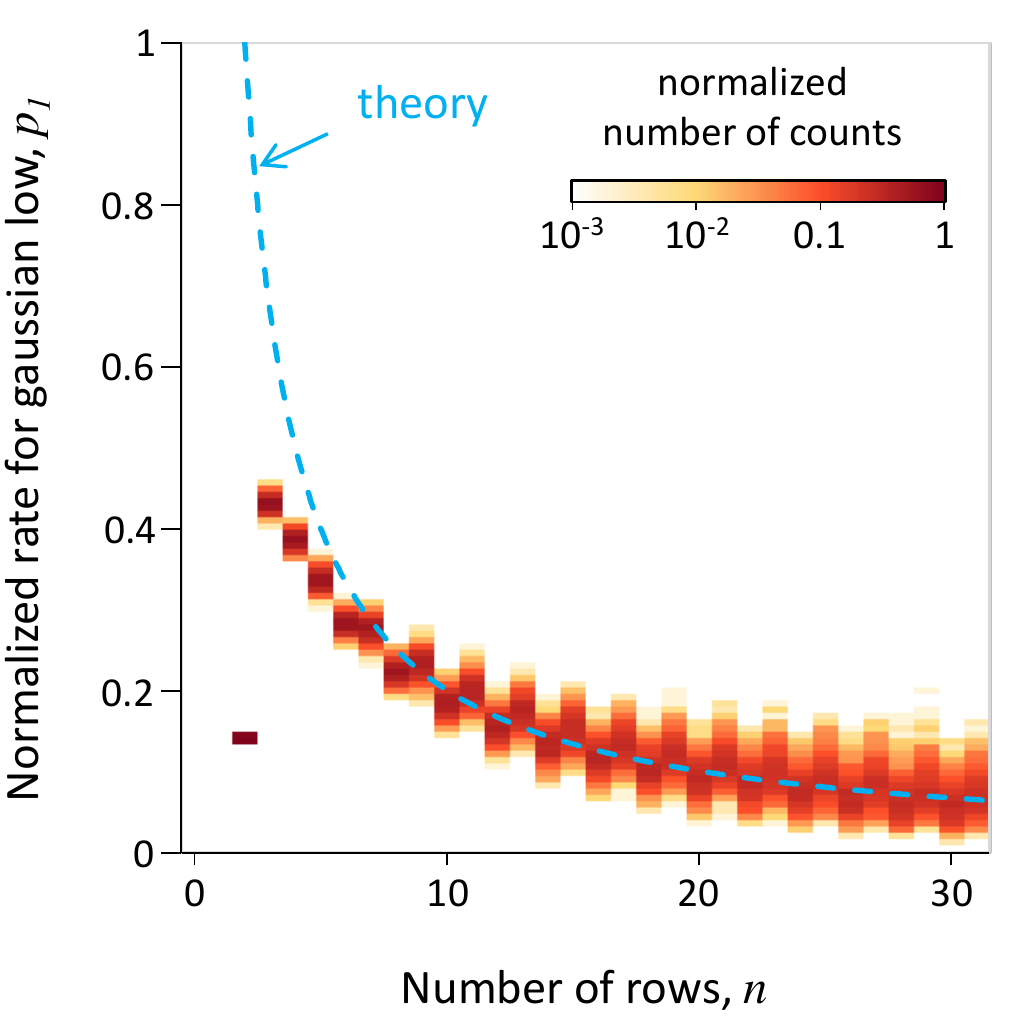}
	\caption{Indirect measurement applied as a way to complete theory (Eq.~(\ref{u_bm})) for the Galton Board. The distribution of reconstructed values of $p_1$ (defines the width of Gaussian distribution) is shown as a function of number of rows $n$. In the limit of large $n$ the distribution show systematic tendency towards the known analytical answer (Eq.~(\ref{a_bm})). The deviation in the region of small $n$ provides an extension of the theory in the framework of Eq.~(\ref{u_bm}).}
	\label{bm_id}
\end{figure} 

As an illustrative example, we again use the Galton Board experiment. We assume that we determined only the fact of having a Gaussian distribution, but did not determine the coefficient that defines the spread. We can use Eq.~\ref{u_bm} with $p_2 = 0$. Then the unknown free parameter is $p_1$ and our purpose is to determine its dependency on the number of rows $n$.

We perform numerical experiments with different number of rows from $n = 2$ to $n = 32$. In all experiments we had $10^5$ beads. Using the same NN trained with noise as in the previous subsection, we reconstruct the value of $p_1$ as a function of $n$. The result shown in Fig.~\ref{bm_id} clearly demonstrates that we can systematically determine the value of $p_1$. As expected, the result is close to the analytical dependency determined by the Eq.~(\ref{a_bm}) and shown with the dashed blue curve. 

Note that for small values of $n$ the determined values show a systematic deviation from the analytical trend. This is an indication that the analytical trend is not valid for small $n$. This procedure provides the possibility to not only see this but also measure and determine the closest choice of $p_1$ in the framework of the theory (Eq.~(\ref{u_bm})). In such a way, based on experimental data, we can perform an indirect measurement of any parameter and its dependency even beyond the range where the parameter has intended physical meaning.

\subsection*{Indirect measurement}

We would like to note that the procedure described in the previous subsection can be applied also for indirect measurement of a real physical variable that defines physical conditions modeled by the theory. This means that we can use this procedure to perform indirect measurements of unknown parameters in the experiments and, in such a way, complete the information about experimental conditions in case they are not known for us. Moreover, instead of a theory we can use ab-initio simulations, which makes the method applicable for many complex situations. To do so one needs to parametrize the conditions of the process using knowledge about real conditions and perform sufficiently large set of ab-initio simulations to train NN for the described procedure.

\subsection*{Why using NN?}

One can ask a very reasonable question: What is the benefit of using a NN instead of collecting all possible outcomes of a model and then determining, which one is the closest to the experimental measurement? To clarify the benefit, we note that this alternative procedure would inevitable require setting some metric for calculating the closeness between the known output and the measured result. This is the main trouble. First, the appropriate metric can be different for different regions in the parameters space. Second, it can be sensitive to the experimental noise. Finally, it can be sensitive to systematic distortions in the experimental measurements. For example, a systematic shift of a certain useful pattern can hamper its identification in case of using the mean standard deviation as the metric. 

In the discussed methods (on the basis of NN), we only set the metric for a number of physical parameters and the effect of the used metric is more transparent. The NN is automatically adjusted to relate the parameters to the most indicative features in the output. As we showed, one can even train NN to have tolerance to noise and some distortions.

\section*{Application for the physics of laser-plasma interactions}

In this section we show how the discussed methodology can be applied for resolving long-standing questions in the field of high-intensity laser-plasma interactions. The process of such interactions is crucial for many applied and fundamental research direction related to the use of modern high-intensity lasers \cite{mourou.rmp.2006}. Even compact table-top lasers can now produce pulses with relativistic intensity, which means that not only almost instantaneous ionization but also a relativistic, collective dynamics in the produced plasma can be caused on the surface of a target. This opens opportunities for driving large variety of highly non-linear interaction regimes and thus for converting laser energy into energetic particles or unique, tailored forms of radiation. This hold promise for numerous applications ranging from fundamental studies to new diagnostic tools in medicine and nuclear waste utilization \cite{esarey.rmp.2009, teubner.rmp.2009, daido.rpp.2012, macchi.rmp.2013}.

The ionization of solids leads to the formation of high-density plasma that hampers the penetration of laser radiation. However, the light pressure can be strong enough to cause relativistic, repeated shifts of electron bulks that is balanced by the attraction to the residual ions that are less mobile due to higher mass. Using the appropriate reference frame and some reasonable assumptions, one can reduce the problem to 1D radiation-plasma dynamics \cite{bourdier.pf.1983}. However, the application of first principles leads to a highly complex problem formulation, which is largely inaccessible for analytical tools. At the same time, the use of ab-initio simulations lacks generality and is also of limited use due to commonly incomplete knowledge about experimental conditions. This naturally impedes search for useful regimes in a multi-dimensional space of parameters. 

One way of overcoming this difficulty is developing phenomenological models \cite{gordienko.prl.2004, pirozhkov.pop.2006, quere.prl.2006, sanz.pre.2012, debayle.pop.2013, debayle.pre.2015}, i.e. theories that are based on introducing entities (and the rules of their behavior) that model patterns systematically observed in ab-initio simulations. The applicability of such models is typically motivated and analyzed based on theoretical estimates. 

Historically the first and probably the most known phenomenological model for the nonlinear radiation reflection from plasmas is referred to as the \textit{relativistic oscillating mirror} (ROM) \cite{gordienko.prl.2004}. The underlying principle of this model states that the reflection happens according to the Leontovich boundary condition (equality of incoming and outgoing energy fluxes) at some oscillating \textit{apparent reflection point}, which can approach but not reach the relativistic speed limit just as real particles. When this point moves against the incident radiation with relativistic velocity, a quick transition is formed in the reflected radiation. High harmonics generated in such a way undergo a universal spectral law $I_k \sim k^{-8/3}$, where $I_k$ is the intensity of $k$-th harmonic \cite{baeva.pre.2006}. Although some other trends in spectra are also observed in simulations \cite{debayle.pre.2015, pirozhkov.pop.2006, boyd.prl.2008, debayle.pop.2013, boyd.pla.2016}, the observation of this trend in some experiments \cite{dromey.natp.2006} established a conviction in the validity and applicability of the ROM model. 

An alternative model is referred to as the \textit{relativistic electronic spring} (RES) \cite{gonoskov.pre.2011, gonoskov.pop.2018}. The underlying principle of this model states that under the light pressure some varied part of foremost electrons becomes and remains bunched, while the resulted bunch moves so that its radiation precisely cancels out the incident radiation in the plasma bulk. The resulted description agrees well with ab-initio simulations in many aspects in a wide range of conditions and provides several important predictions \cite{serebryakov.pop.2015, svedungwettervik.pop.2018} including the possibility of producing unprecedentedly intense and short bursts \cite{gonoskov.pre.2011, bashinov.epjst.2014, fuchs.epjst.2014} of radiation with controllable ellipticity \cite{blanco.pop.2018}. However, the experimental validation of the RES model is difficult. Current experiments are largely limited to the observation of high-harmonic spectra, which are typically analyzed in terms of the exponent of the power-law fall, while setting aside more peculiar signatures that can indicate the validity and applicability of the RES model. More accurate comparison requires cutting-edge experimental and theoretical developments based on reveling and retrieving indicative features in the experimental data \cite{thaury.natp.2007, vincenti.natc.2014, borot.natp.2012, kormin.natcom.2018}. We now show how such analysis can be performed with the help of a NN on the basis of the methodology described in the previous section.

\subsection*{Comparing the RES and the ROM models}

Here we show how we can use the measurable spectra of generated high-harmonics for comparing and determining the validity regions of the RES and ROM models. Although the procedure can be based on experimental data, here we use ab-initio simulations to obtain the spectra in the frequency range of up to harmonic order 12.8, which mimics the capabilities of typical experimental arrangements.

To perform the comparison, we need to develop a unification theory. Since the ROM model was intended and motivated for the case of sharp density drop at the plasma surface, we consider the RES equations for this case and also assume the most indicative case of P-polarized incidence:
\begin{eqnarray}
&& {f_{\text{in}}\left(x_s - t\right) = \frac{S}{2\cos^3\theta} \left(\sin\theta - \frac{\beta_y}{1 - \beta_x}\right),}\label{res:1}\\
&& {\beta_x^2 + \beta_y^2 = 1},\label{res:2}\\
&& {\frac{dx_s}{dt} = \beta_x},\label{res:3}\\
&& {f_{\text{out}}^{RES}\left(x_s + t\right) := - \frac{S}{2\cos^3\theta} \left(\sin\theta - \frac{\beta_y}{1 + \beta_x}\right).\label{res:4}}
\end{eqnarray}

Here $f_{\text{in}}$ and $f_{\text{out}}^{RES}$ are the shape of the incoming laser pulse and the shape of the outgoing reflected signal that carries generated high harmonics. 
The first three equations can be solved to find the temporal evolution $x_s(t)$ of the bunch that accommodates peripheral electrons according to the RES model. We can then substitute the obtained solution $x_s(t)$ into the fourth equation to obtain the outgoing signal $f_{\text{out}}^{RES}(x+t)$. The second equation implies the ultra-realistic limit for the bunch velocity components along the pulse propagation direction ($\beta_x$), and along the electric field direction ($\beta_y$). The shapes $f_{\text{in}}$ and $f_{\text{out}}^{RES}$ are given in laboratory reference frame, while the consideration is carried out in the moving reference frame\cite{bourdier.pf.1983} that provides a way to account for arbitrary incidence angle $\theta$. The relativistic similarity parameter $S$ is defined as $S = n/a$, where $n$ is the plasma density given in laboratory reference frame in critical units, $a$ is the pulse field amplitude given in relativistic units; both units are computed relative to the laser wavelength $\lambda = 1$~$\mu$m. For more details see Ref.~\cite{gonoskov.pre.2011, gonoskov.pop.2018}.

The ROM model does not provide a complete set of differential equations for computing $f_{\text{out}}^{ROM}$. Instead the model provides a way to compute directly the indicative spectral properties. The only two essential assumptions that lead to this result are the Leontovich boundary conditions ($f_{\text{out}}^{ROM}(x_{ARP} + t) = - f_{\text{in}}(x_{ARP} - t)$) and the fact that they are applied to the point $x_{ARP}$ that passes through the stage of moving with the speed close to the speed of light against the incident radiation (in the moving reference frame). From Eqs.~(\ref{res:1}) and (\ref{res:4}) we see that the boundary conditions of the RES model implies inequality between the incident and outgoing radiation. To provide a smooth transition between these types of boundary conditions and use $x_s$ as $x_{ARP}$ (i.e. admitting the same relativistic motion) we modify the Eqs.~(\ref{res:1} - \ref{res:4}) and formulate the unification theory:

\begin{eqnarray}
&& {f_{\text{in}}\left(x_s - t\right) = \frac{S}{2\cos^3\theta} \left(\sin\theta - \frac{\beta_y}{\left(1 - \beta_x\right)^{\frac{p+1}{2}}}\right),}\label{u:1}\\
&& {\beta_x^2 + \beta_y^2 = 1},\label{u:2}\\
&& {\frac{dx_s}{dt} = \beta_x},\label{u:3}\\
&& {f_{\text{out}}^{U}\left(x_s + t\right) := } \nonumber\\
&& {- \frac{S}{2\cos^3\theta} \left(\sin\theta - \frac{\beta_y}{\left(1 - \beta_x\right)^{\frac{1-p}{2}} \left(1 + \beta_x\right)^p}\right).\label{u:4}}
\end{eqnarray}

Here we introduce the parameter $p$ that provides the needed transition through its variation from 0 to 1. In the limit $p = 1$ the Eqs.~(\ref{u:1} - \ref{u:4}) coincide exactly with the equations of the RES model. In the limit $p = 0$ the Eqs.~(\ref{u:1}) and (\ref{u:4}) imply the Leontovich boundary conditions, while the solutions for $x_s$ include instances of approaching relativistic limit of motion against the incident radiation (see more details below). Thus the unified theory for $p = 0$ exposes the same spectral properties as the ROM model. However, the unified theory still provides one particular way of completing the ROM model to a set of equations that determine not only spectral properties, but also the explicit form of $f_{\text{out}}$. We thus will refer to this theory as the completed ROM or ROM$^\text{c}$.

The numerical solution of Eqs.~(\ref{u:1} - \ref{u:3}) is straightforward. From Eq.~(\ref{u:1}) one can express:
\begin{equation}\label{rr}
\frac{\beta_y}{\left(1 - \beta_x\right)^{\frac{p+1}{2}}} = R = \sin\theta - 2 S^{-1}\cos^3\theta f_{\text{in}}\left(x_s - t\right).
\end{equation}
One can see that when $R$ changes sign, so does $\beta_y$.  According to Eq.~(\ref{u:2}), this means that $\beta_x$ passes close to 1 or -1 as it was mentioned above. The Eq.~(\ref{rr}) together with Eq.~(\ref{u:2}) have one relevant solution for $\beta_x$:
\begin{equation}
\beta_x = 1 - g\left(R^2, p\right),
\end{equation}
where $g\left(\alpha, p\right)$ is the indirect solution of the equation $\alpha x^p + x - 2 = 0$. We can solve this equation numerically and use this to determine the evolution $x_s(t)$. Then we can use Eq.~(\ref{u:2}) to obtain $f_{\text{out}}^{U}\left(x_s + t\right)$ and calculate its spectrum.

Once the unification theory is developed we can use it to calculate the spectra. For our studies we considered two-cycle laser pulse $f_{\text{in}}\left(\eta = x - t\right)$ characterized by the vector potential in the form of $\sim \sin\left(\eta + \phi\right) \sin^4\left(\eta/4\right)$, which results in:
\begin{eqnarray}
&& {f_{\text{in}}\left(\eta = x - t\right) = \sin^3\left(\eta/4\right) \times} \nonumber \\
&& {\left(\cos\left(\eta/4\right)\sin\left(\eta + \phi\right) + \sin\left(\eta/4\right)\cos\left(\eta + \phi\right)\right)}, \label{pulse}
\end{eqnarray}
where the phase $\phi$ can have arbitrary value. For this study we set $\phi = \pi/2$.

We numerically solve the equations of the unified theory for the parameter space spanned by:
\begin{eqnarray}
&& {\theta \in \left[0, 3 \pi/8\right],}\label{p:1}\\
&& {S \in \left[1, 10\right]},\label{p:2}\\
&& {p \in \left[0, 1\right]}.\label{p:3}
\end{eqnarray}
Each spectrum was sampled with 16 equidistant points in the interval from 0 to harmonic order 12.8 and the value is converted to appropriate logarithmic units so that the values mostly lie in the interval from 0 to 1. Next we train the NN to reconstruct the values of $S$, $\theta$ and $p$ from this data. We use the same topology and training method of the NN as in the previous experiments with the Galton Board. The achieved accuracy was about $3 \times 10^{-3}$ for the square error measure applied to the parameters normalized to unity.

Next, we perform a series of particle-in-cell (PIC) simulations for the parameter space spanned by Eqs.~(\ref{p:1} - \ref{p:2}) and obtain spectra using the field distribution obtained in each simulation. For this purpose we used 1D version of PIC code ELMIS \cite{gonoskov.2013, gonoskov.pre.2015}. In all simulations we used field amplitude $a = 200$ and the density determined in accordance to the value of $S$. The spectra are then sent to the input of the trained NN that provides the reconstructed values of parameters.

\begin{figure}[t!]
	\centering\includegraphics[width=0.5\columnwidth]{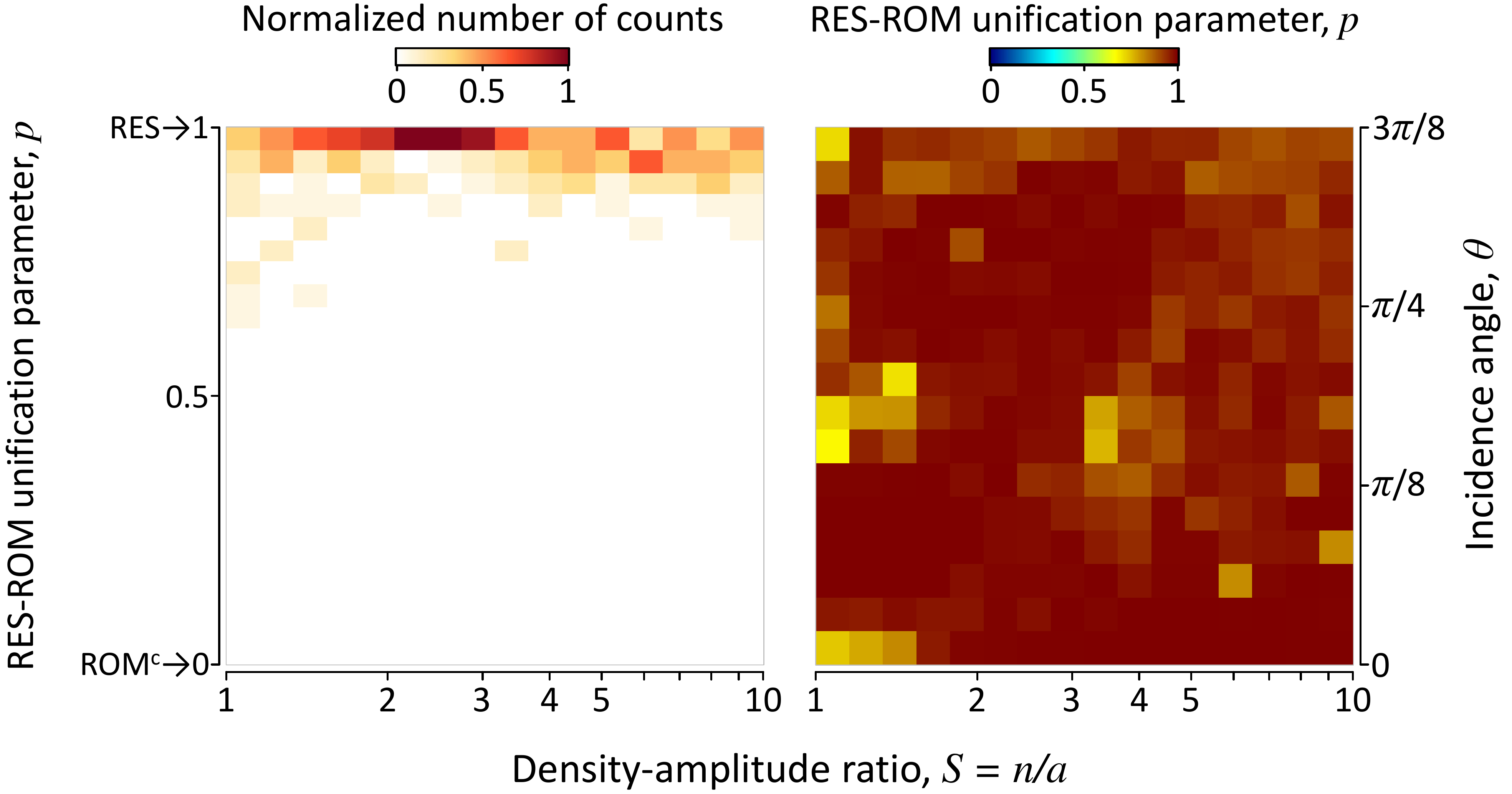}
	\caption{Comparison of the RES and the ROM model. The RES-ROM unification parameter $p$ (see Eqs.~(\ref{u:1} - \ref{u:3})) is shown as a function of the incidence angle $\theta$ and $S$ (right panel). The distribution of $p$ as a function of $S$ is shown on the left panel.}
	\label{res_tv}
\end{figure} 

In Fig.~\ref{res_tv} we show the distribution of the values of the parameter $p$ as a function of $S$ and $\theta$. The results point to the fact that in all cases the most appropriate choice of $p$ is close to 1, which corresponds to the RES model.

\subsection*{Advancing the RES model}

The original RES model has no fitting parameters and still describes fairly accurately the shape of the pulse computed with ab-initio PIC simulations for a wide range of relativistic laser-plasma interaction scenarios \cite{gonoskov.pop.2018, blackburn.pra.2018}. In particular, the model indicates the possibility of producing singularly intense XUV bursts, identifies the optimal conditions for this process \cite{gonoskov.pre.2011, blackburn.pra.2018} and determines the polarization states of the XUV bursts \cite{blanco.pop.2018}. However, the original RES theory does not predict the amplitude of these bursts that appear as singular points for the theory.

These bursts are originated from the singularity of the second term in the right-hand side of Eq.~(\ref{res:4}), when $\beta_y$ changes sign and $\beta_x$ becomes close to -1 according to Eq.~(\ref{res:2}). To see this one can substitute $\beta_y/(1 - \beta_x)$ expressed from Eq.~(\ref{res:1}) into Eq.~(\ref{res:4}):
\begin{equation}
f_{\text{out}}^{RES} := -\frac{S\sin\theta}{\cos^3\theta} + f_{\text{in}} + \frac{2}{1 + \beta_x} \left( \frac{S\sin\theta}{2\cos^3\theta} - f_{\text{in}} \right).\\
\end{equation}
One can formally bound the resulting field through introducing a constant $\alpha$ that is close to, but less than, unity:
\begin{equation}\label{res:alpha}
f_{\text{out}}^{RES} := -\frac{S\sin\theta}{\cos^3\theta} + f_{\text{in}} + \frac{2}{1 + \alpha\beta_x} \left( \frac{S\sin\theta}{2\cos^3\theta} - f_{\text{in}} \right),\\
\end{equation}
and also relate $\alpha$ to the effective bounding gamma factor $\gamma_b$ in terms of Eq.~(\ref{res:2}):
\begin{equation}
\alpha = \sqrt{1 - \gamma_b^{-2}}.
\end{equation}
Large values $\gamma_b > 10$ result in $0.99 < \alpha < 1$ and thus almost do not affect the values of $f_{\text{out}}^{RES}$ everywhere except the vicinity of $\beta_x = -1$, where they formally bound the result. This bound, however, affects crucially the high-frequency end of the spectra measurable in experiments. Simulations show \cite{bashinov.epjst.2014} that the amplitude of the XUV bursts can be up to factor 20 times higher than that of the incident radiation and this factor grows with the laser intensity in a complex way. This means that determining $\gamma_b$ is a matter of theory beyond the self-similarity $S = n/a$ implied by the RES model. Determining $\gamma_b$ is thus an important theoretical problem for both experimental validations and future applications of ultra-intense XUV bursts.

Serebryakov et al. (see Ref.~\cite{serebryakov.pop.2015}) analyzed the possibility to relate the bounding factor to the actual gamma factor of electrons in the bunch. However, the electrons in the bunch have only similar velocity, but genuinely different gamma factors, since they all experience different acceleration over different intervals of time \cite{gonoskov.pop.2018}. 

We will now use the methodology of indirect measurements to examine heuristically the appropriate values of $\gamma_b$ as a function of laser field amplitude. We use the RES model extended with parameter $\alpha$ through Eq.~(\ref{res:alpha}). We obtain the same spectral data for various parameters in the space spanned by:
\begin{eqnarray}
&& {\theta \in \left[0, 3 \pi/8\right],}\label{p2:1}\\
&& {S \in \left[1, 10\right]},\label{p2:2}\\
&& {\alpha \in \left[0, 1\right]}.\label{p2:3}
\end{eqnarray}
We train the same kind of NN and reach roughly the same level of accuracy of determining these parameters on the basis of spectral data. 

\begin{figure}[t!]
	\centering\includegraphics[width=0.5\columnwidth]{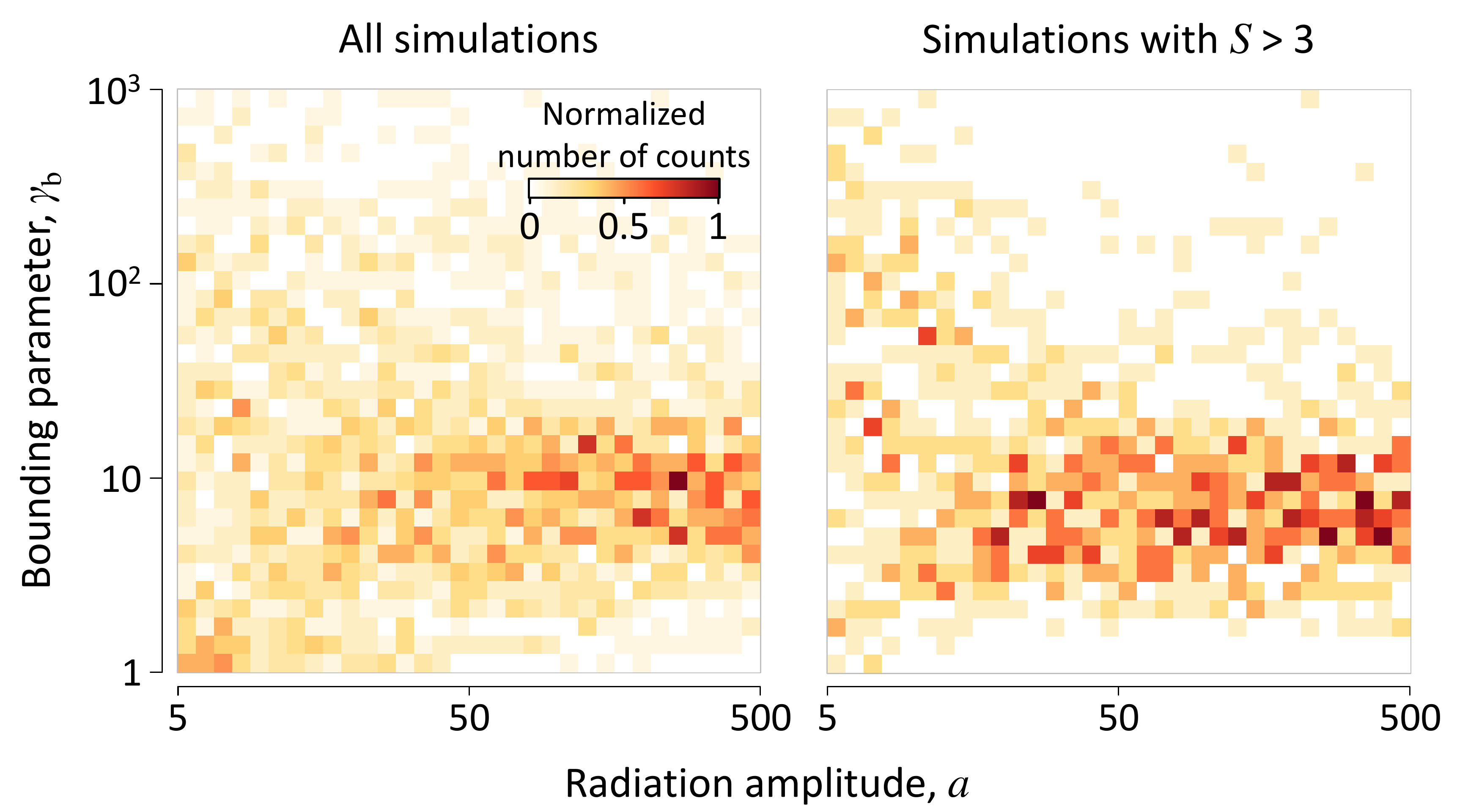}
	\caption{Indirect measurement of the $\gamma_b$ parameter for the extension of the RES model according to Eq.~(\ref{res:alpha}). The distribution of reconstructed values of $\gamma_b$ are shown as a function of the laser field amplitude $a$ for $1 < S < 10$ (left panel) and for $3 < S < 10$ (right panel).}
	\label{res_gd}
\end{figure} 

Next, we perform a series of PIC simulations for $\theta \in \left[0, 3 \pi/8\right]$, $S \in \left[1, 10\right]$ and field amplitudes $a \in \left[5, 500\right]$, which is relevant to current and near-future experiments. The obtained spectra are used as inputs for the trained ANN that reconstructs the values of $\alpha$. In Fig.~\ref{res_gd} we see that heuristic value of $\gamma_b \approx 10$ appears as universal for the amplitude values $a \gtrsim 10$. On the right panel we also see that this tendency becomes more prominent for large values of $S$. 

This result can be used either directly to advance heuristically the RES model or for further theoretical analysis of the physics of the process. Note, that since the parameter $\gamma_b$ has no meaning in terms of first principles we can hardly determine it from these principles. In opposite, we here determine its value in terms of its intended role as a parameter of the extended RES model.

\subsection*{Indirect measurements based on incomplete knowledge about experimental conditions}

The previous examples primarily concerned the questions of theory. As a final example we demonstrate how we can make a clear use of the discussed methodology in complex experiments, when the information about the initial conditions of the studied process is not complete. This is not related to the use of any models and can be based on ab-initio simulations. However, here, we again use the RES model and ab-initio PIC simulations to mimic such experimental scenarios.

Suppose we perform an experiment about high-harmonic generation through the interaction of a high-intensity laser pulse with a solid target. We know and can vary the incidence angle $\theta$. We also know the duration and the amplitude of the laser pulse. However, we do not know the carrier envelope phase (CEP) $\phi$ and we do not know precisely the density profile that is the result of plasma spreading after heating by foregoing laser radiation. This is rather common experimental situation. 

We will mimic the unknown initial plasma state through considering a steep density profile with unknown density in the plasma bulk. This can be related to any plasma distribution through the approach of effective $S$-number proposed in Ref.~\cite{blackburn.pra.2018}. Note, however, that this is just for showing proofs of principles and one can use any plasma density profile in both RES and PIC calculations \cite{gonoskov.pop.2018}. 

We will assume that the pulse has the form given by Eq.~(\ref{pulse}) with amplitude $a = 200$. The limited knowledge about plasma density $n$ can be interpreted as limited knowledge about the $S = n/a$, which we assume to stay in the range $S \in \left[1, 10\right]$.

\begin{figure}[t!]
	\centering\includegraphics[width=0.5\columnwidth]{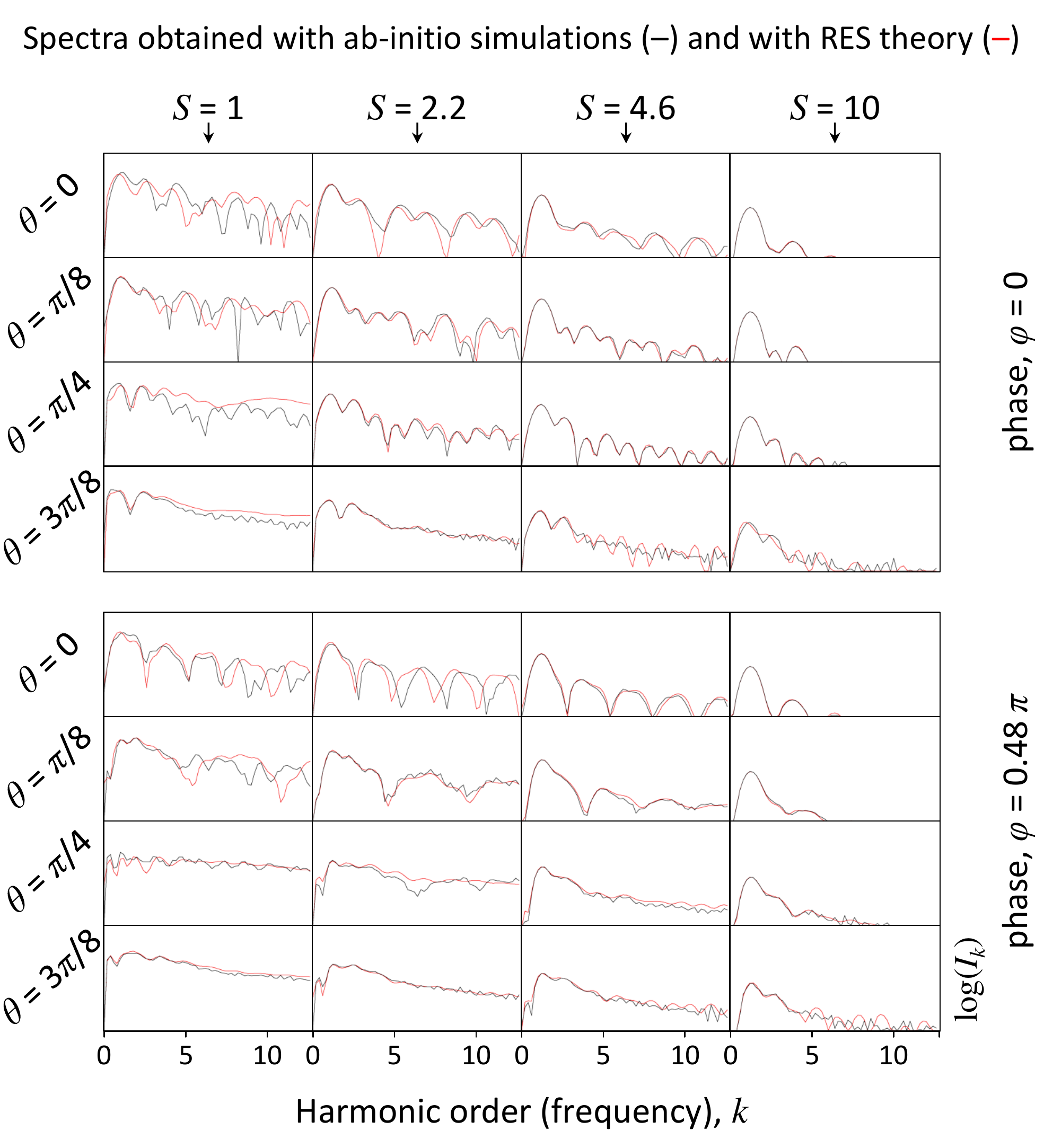}
	\caption{Some spectra obtained with the PIC simulations (black) and the numerical solution of the RES model (red) for different parameters $S$, $\theta$ and $\phi$.}
	\label{res_ser}
\end{figure} 

In Fig.~\ref{res_ser} we show common spectral data obtained for various values of $S$, $\theta$ and $\phi$. One can see that the data contains sophisticated features that depend on parameters in a complex way. Although they seemed to encode ambitiously the scenario of interaction and the initial conditions, it would be very difficult to describe them using human language so that one can determine the initial conditions systematically. Developing a methodology of retrieving such information from the spectral data appears as an intrinsically complex problem that is a matter of advanced developments \cite{borot.natp.2012, kormin.natcom.2018}. Here we demonstrate that this problem can be solved with a NN.

Note, that in essence the results of the RES model (red curves) agree with the results of PIC simulations. However, since the agreement is not ideal and in some cases is rather poor, the problem of reconstructing initial conditions based on the RES model is not trivial. This requires appealing to some essential features rather than to ideal memory about the states. This mimics the possible experimental limitations related to natural noise or potentially systematic distortions.

\begin{figure}[t!]
	\centering\includegraphics[width=0.5\columnwidth]{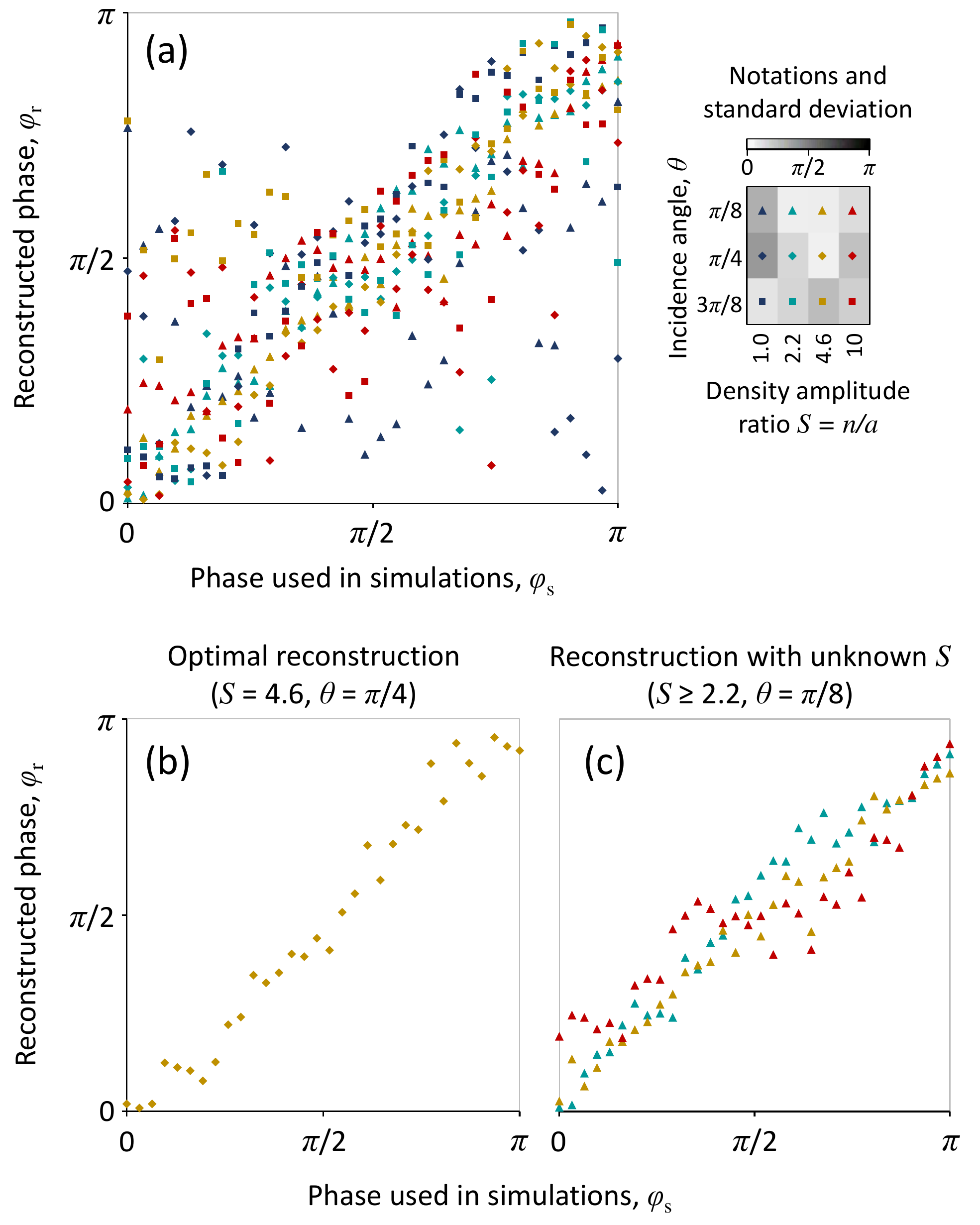}
	\caption{The demonstration of indirect measurement of the pulse carrier envelope phase $\phi$ based on spectral data obtained with PIC simulations. The used NN was trained with the RES model. Panel (a) shows all results and the insert to the left shows the notations and standard deviation for different groups of PIC simulations. Panel (b) shows the results for $\theta = \pi/4$ and $S = 4.6$ that are optimal for accurate reconstruction. Panel (c) shows the result for the optimal $\theta = \pi/8$ in case we have no precise information about $S \in \left[2.2, 10\right]$.}
	\label{res_ph}
\end{figure}

We use the same kind of NN and train it to reconstruct values of $\phi$, $\theta$ and $S$ on the basis of spectral data obtained via numerical solutions of the RES model in the parameter space spanned by:
\begin{eqnarray}
&& {\theta \in \left[\pi/8, 3 \pi/8\right],}\label{p3:1}\\
&& {S \in \left[1, 10\right]},\label{p3:2}\\
&& {\phi \in \left[0, \pi\right]}.\label{p3:3}
\end{eqnarray}
To mimic real experiments we perform PIC simulations with the parameters in the same parameter space. We then use the obtained spectral data to see whether the NN can identify correctly the phase values used in simulations. The results are shown in Fig.~\ref{res_ph}.

As one can see in Fig.~\ref{res_ph}~(a), all the results of the NN are mostly located around the diagonal that corresponds to accurate reconstruction of the phase $\phi$. However, from the diagram for the standard deviation (shown in the insert to the right) the accuracy of the reconstruction varies with parameters $\theta$ and $S$. 

There could be two reasons for this. First, potentially the RES model has different accuracy for different parameters and this affects the accuracy of reconstruction. Second, the physics of the process can potentially provide more indicative feature for certain ranges of parameters. This conclusion indicates an important capability of this methodology. The accuracy of the reconstruction of the known parameters can show where (in the parameter space) the used theoretical model (or the setup for simulations) is more adequate. Alternatively, this accuracy can indicate where the physics of the process provides more indicative features in the data used for the reconstruction. Moreover, if we can vary the range of used data, we can determine where these features are located.

One trivial outcome of this observation is the possibility to chose the parameters that are most useful for the reconstruction. In our case this is show in Fig.~\ref{res_ph}~(b).

Finally, we demonstrate that the procedure can be efficiently used in case of limited knowledge about the experimental data. As we see in Fig.~\ref{res_ph}~(c), the phase can be reconstructed fairly accurate even if we do not know precisely the plasma density distribution.

\section*{Conclusions}

In this paper we discussed and demonstrated the possibility of using machine learning for validating and advancing theories, as well as performing indirect measurements with incomplete knowledge about experimental conditions. The procedure is based on the possibility of using NN for establishing the relation between various parameters (of the process and theory) and the features that might be poorly accessible for description with human language. First, we showed how this can be used to validate, compare and advance theoretical models. Next, we showed how this can be used to perform indirect measurement of parameters of the experiment or theory based on experimental data, even if we have incomplete knowledge about experimental conditions. Finally, we outline that one can use the accuracy of the reconstruction of the known parameters for the identification of indicative features and their locations in the experimental data.

\normalem

\bibliography{fml}

\section*{Acknowledgements}

A.G. would like to thank M. Marklund and T. G. Blackburn for useful discussions. Simulations were performed on resources provided by the Swedish National Infrastructure for Computing (SNIC) at the High Performance Computing Centre North (HPC2N). The research was supported by the Swedish Research Council under Grant No. 2017-05148.

\section*{Author contributions statement}

All authors contributed to discussions and preparation of the manuscript, A.~P. and I.~M. were responsible for machine learning techniques, E.~W. performed some of the numerical experiments, A.~G. proposed the methodology and led the project.\\

\section*{Additional information}

\textbf{Competing Interests:} The authors declare that they have no competing interests.

\end{document}